# Emergent Dynamic Magnetic Ground State in a Mixed 3d/5d Heavy Fermion System $CaCu_3Ir_4O_{12}$

J. Ming[1, +, #], Abhisek Bandyopadhyay[1,2, *, #], G. B. G. Stenning[1], M. T. F. Telling[1], N. N. Wang[3,4], G. Wang[3,4], J.-G. Cheng[3,4] and D. T. Adroja[1,5, $]

[1]ISIS Neutron and Muon Source, STFC, Rutherford Appleton Laboratory, Chilton, Oxon OX11 0QX, United Kingdom

[2] Department of Physics, Ramashray Baleshwar College (A Constituent Unit of Lalit Narayan Mithila University, Darbhanga, India), Samastipur, Bihar 848114, India

[3]Beijing National Laboratory for Condensed Matter Physics, and Institute

of Physics, Chinese Academy of Sciences, Beijing 100190, China.

[4]School of Physical Sciences, University of Chinese Academy of Sciences, Beijing 100190, China.

[5]Highly Correlated Matter Research Group, Physics Department, University of Johannesburg, Auckland Park 2006, South Africa

[+]jingming1017@gmail.com

[*]abhisek.ban2011@gmail.com

[$]devashibhai.adroja@stfc.ac.uk

[#]J.M. and A.B. equally contribute to this work

# Abstract

Quantum-disordered magnetic ground states are challenging to identify in three-dimensional (3D) oxides, where strong exchange pathways typically favour long-range magnetic order or spin freezing. The quadruple perovskite $CaCu_3Ir_4O_{12}$, crystallizing in the cubic *Im-3* structure, provides a 3D lattice where $Cu^{2+}$ 3d-moments are coupled to an extended Ir 5d network, offering a rare platform for probing quantum-disordered magnetism in a mixed 3d/5d electron system. Here, we combine bulk probes, including DC, AC magnetic susceptibility, and heat capacity (down to 50 mK), along with the local microscopic probe muon spin relaxation (μSR) (down to 40 mK) to investigate the true magnetic ground state of $CaCu_3Ir_4O_{12}$. Despite strong antiferromagnetic interactions ($\theta_W$ ~ −200 K, which has applied field dependency), no signature of long-range magnetic ordering or spin freezing is detected down to the lowest measured temperatures. Further, our in-depth zero-field (ZF) and longitudinal-field (LF) μSR characterizations confirm strong quantum spin fluctuations and truly dynamic nature of the local moments till down 40 mK. These results establish $CaCu_3Ir_4O_{12}$ as a promising 3D quantum-disordered magnet and a well-characterized platform for exploring fluctuation-dominated states in correlated 3d/5d oxides.

## 1. Introduction

Quantum spin liquids (QSLs) are intriguing magnetic states in which strong quantum fluctuations prevent conventional symmetry-breaking order, even as the temperature approaches absolute zero. In contrast to most conventional magnets, which evolve into statically ordered states, frustrated systems may host quantum-disordered phases that are closely linked to topological phases of matter [1,2]. Both two-dimensional (2D) and three-dimensional (3D) systems provide routes to QSL behaviour: low-dimensional lattices generally favour realization of QSL states, whereas 3D systems offer the possibility of intrinsically more robust quantum-disordered states [2,3]. One established route to QSL behaviour arises from geometric frustration, as in the landmark case of the hyperkagome iridate $Na_4Ir_3O_8$, in which a 3D frustrated $Ir^{4+}$ network avoids long-range magnetic order down to 2 K [4,5]. Beyond geometric frustration, it has been recognized that strong spin–orbit coupling (SOC) in heavier 4d and 5d transition-metal compounds further enriches this landscape by generating highly anisotropic, bond-dependent exchange interactions that can suppress classical magnetic order even in structurally ordered lattices, as proposed in Kitaev materials such as the honeycomb iridate $Na_2IrO_3$ [3,6,7]. This motivates the search for chemically ordered three-dimensional iridium-based oxides as compelling candidates for quantum-disordered magnetism, in which the interplay between SOC and competing exchange interactions can be tuned in a controlled manner.

Among such 3D candidate platforms, the family of *A*-site-ordered quadruple perovskites, $(AA'_3)B_4O_{12}$, has attracted growing attention as a chemically ordered, fully 3D variant of the simple perovskite structure. In these systems, the *A* and transition-metal *A′* sites are crystallographically ordered on distinct sublattices, while the *B*-site cations form a three-dimensional network of corner-sharing $BO_6$ octahedra. Substitution at the *B* site over a wide range of 3*d*, 4*d*, and 5*d* transition-metal ions enables systematic tuning of electronic correlations and magnetic properties while preserving the cubic lattice symmetry ($Im\bar{3}$) [8]. Within this class, compounds with Cu occupying the *A′* site feature square-planar $CuO_4$ units embedded in a 3D network of corner-sharing $BO_6$ octahedra. This geometry allows direct coupling of localized Cu-derived moments to the extended *B*-O framework, providing a well-ordered platform to study spin–orbit–driven magnetic physics (Fig 1). A prominent example is $CaCu_3Ru_4O_{12}$, which has long been discussed as a *d*-electron heavy-fermion-like material, illustrating how coupling between Cu-derived localized moments and itinerant Ru-4*d* electrons can lead to anomalous low-temperature behaviour [8–12]. Motivated by this interplay, here, we focus on the Ir-analogue $CaCu_3Ir_4O_{12}$, where $Cu^{2+}$ moments coexist with an extended 5d framework of $Ir^{4+}$, bringing strong SOC, electron correlation, and 3*d*-5*d* hybridization into a single ordered framework. Early work on $CaCu_3Ir_4O_{12}$ suggested Kondo-like behaviour near a metal–insulator crossover, associated with coupling between localized Cu 3*d* orbitals and itinerant Ir–O bands, and identified a characteristic anomaly near ~80 K [13]. This feature was reproduced within a two-fluid model analysis of the magnetic susceptibility, invoking coexisting Fermi-liquid and heavy-fermion-like components of the Ir $d_{xy}$ electrons below this temperature [13].

Whether $CaCu_3Ir_4O_{12}$ hosts a genuine quantum-disordered ground state is, however, far from straightforward to establish, since identifying such states in three-dimensional materials remains particularly challenging. Despite substantial progress in the field, identifying QSLs remains inherently puzzling because many candidate materials ultimately develop long-range order due to weak perturbations, or exhibit glassy freezing that can mimic “disordered” magnetism without representing a genuine spin-liquid ground state [14]. These challenges are amplified in 3D oxides, where the larger coordination number and additional exchange pathways generally reduce quantum fluctuations and tend to stabilize conventional ordered magnetic phases [2,14,15]. In practice, even in presence of strong magnetic frustration, chemical disorder can further blur the distinction between an intrinsically dynamic QSL and a “QSL-like” response arising from glassiness or inhomogeneous magnetism, as illustrated by site mixing in herbertsmithite [16], random local environments in $YbMgGaO_4$ [17], and more broadly non-stoichiometry or chemical inhomogeneity, oxygen vacancies that introduces Curie tails, freezing, or sample-dependent low-temperature anomalies [18].

Against this backdrop, an important question is whether $CaCu_3Ir_4O_{12}$ develops static magnetic order or spin freezing at low temperature, or instead remains dynamically fluctuating down to the millikelvin regime? Addressing this question is nontrivial, since bulk thermodynamic measurements alone often cannot distinguish weak static internal fields associated with magnetic order or partial spin-freezing from the persistent spin dynamics. Establishing the magnetic ground state therefore requires combining bulk probes with local microscopic techniques such as muon spin relaxation ($\mu$SR), which can directly distinguish static from dynamic magnetism [19]. To clarify the true magnetic ground state of $CaCu_3Ir_4O_{12}$, we combine bulk magnetic and thermodynamic measurements with $\mu$SR to probe its magnetism down to millikelvin temperatures. This integrative approach enables static and dynamic magnetic responses to be disentangled across multiple length scales. Such insight may provide a microscopic basis for understanding quantum-disordered magnetism in chemically ordered three-dimensional iridates and help guide the search for related candidate materials.

## 2. Experimental Methods

The $CaCu_3Ir_4O_{12}$ polycrystalline samples used in this study were prepared following the high-pressure synthesis procedure reported previously [13]. The phase purity was examined by room-temperature powder X-ray diffraction (PXRD) using a Philips X'Pert diffractometer with Cu $K_\alpha$ radiation. Rietveld refinement of the PXRD data was performed using the FullProf software package [20]. The fitted diffraction profiles are shown in Figure 1a, while the crystallographic and refinement parameters are given in Table S1 and S2 in the supplementary materials (SM).

DC magnetic susceptibility, $\chi$, as a function of temperature was measured using a Quantum Design MPMS3 superconducting quantum interference device equipped with a vibrating sample magnetometer (VSM) in the temperature range of 2–300 K and in applied magnetic fields $H$ of up to 50 kOe in both Zero-field-cooled (ZFC) and field-cooled (FC) modes. Further, the magnetization isotherms betweeen 2 and 30 K were measured in applied fields up to ± 60 kOe.

Specific-heat, $C_p$, measurements were performed using a Quantum Design Physical Property Measurement System (PPMS), employing the 2-$\tau$ thermal relaxation method from 2 K to 300 K under magnetic fields up to 90 kOe. To access the low-temperature regime, additional specific heat and AC magnetic susceptibility $\chi_{ac}$ measurements were conducted between 0.05 and 4 K using a Quantum Design DynaCool PPMS equipped with a $^3He/^4He$ dilution insert [21].

Zero-field (ZF) and longitudinal-field (LF) muon spin rotation/relaxation (μSR) experiments were performed on the EMU spectrometer at the ISIS Neutron and Muon Source, UK. Powdered $CaCu_3Ir_4O_{12}$ samples were mounted on a high-purity (99.995%) silver sample holder using GE varnish to provide good thermal coupling. The sample was covered with a thin silver foil and cooled to temperatures down to 0.04 K using a dilution refrigerator. Analysis of the μSR asymmetry spectra was carried out using the Mantid software framework [22].

# 3. Results and Discussion

## 3.1 Crystal Structure

The Rietveld-fitted powder XRD (PXRD) pattern confirms nearly pure single-phase of $CaCu_3Ir_4O_{12}$ in a cubic *A*-site-ordered quadrupole perovskite structure with space group $Im\bar{3}$ (*No*. 204), in agreement with the reference pattern (ICSD-251658) [23]. In addition to the desired primary $CaCu_3Ir_4O_{12}$ phase (~97 wt%), minor proportions of ~1 wt% $IrO_2$ and ~2 wt% Ir-metal as secondary phases are detected in our PXRD refinement, as shown in Fig 1a. The refined crystallographic parameters and the refinement quality factors have been listed down in Table I in the Supplemental Material. The refined atomic positions are also fully consistent with those reported for the other members of the $CaCu_3B_4O_{12}$ family (*B* = Ru, Pt, Ge, Sn, etc.), essentially reflecting the unchanged underlying lattice connectivity across the series [8,24,25]. Such structural similarity therefore suggests that the difference in the low-*T* magnetic behaviours across this quadruple perovskite series primarily arise from electronic effects governed by the *d*-electron occupancy associated with the choice of *B* cation.

In the $(AA'_3)B_4O_{12}$ structure, $Ca^{2+}$ occupies the larger *A* site, while $Cu^{2+}$ resides on the A′ sublattice in square-planar coordination, forming $CuO_4$ plaquettes, with an A : A′ site ratio of 1:3. The Cu–O bond lengths (1.948 and 2.856 Å) confirm square-planar $CuO_4$ coordination (Figs.1b-c), supporting localized $Cu^{2+}$ moments, while the Ir–O distances (1.999 Å) indicate an $IrO_6$ octahedral network forming an extended 5d framework where spin–orbit coupling is significant. The $Im\bar{3}$ quadruple-perovskite structure intrinsically enforces bent Ir–O–Ir (138.4°) (Fig.1d) and Cu–O–Ir (110.1°) exchange pathways rather than linear bonds (Fig.1b) [13]. This geometry implies more effective orbital overlap within the Ir–O network than along the more strongly constrained Cu–O–Ir pathways, pointing to the spin–orbit-coupled Ir framework that is structurally connected to, yet not rigidly locked by, the Cu-derived local moments [26]. The Cu–O–Ir connectivity further provides a coupling channel between localized Cu moments and the extended Ir–O electronic states, weakening the dominance of a single isotropic super-exchange interaction and enhancing the role of competing anisotropic exchange interactions [27]. Together, these structural features confirm the chemically ordered nature of $CaCu_3Ir_4O_{12}$, making it a suitable system for exploring magnetic ground state in a chemically ordered three-dimensional oxide framework.

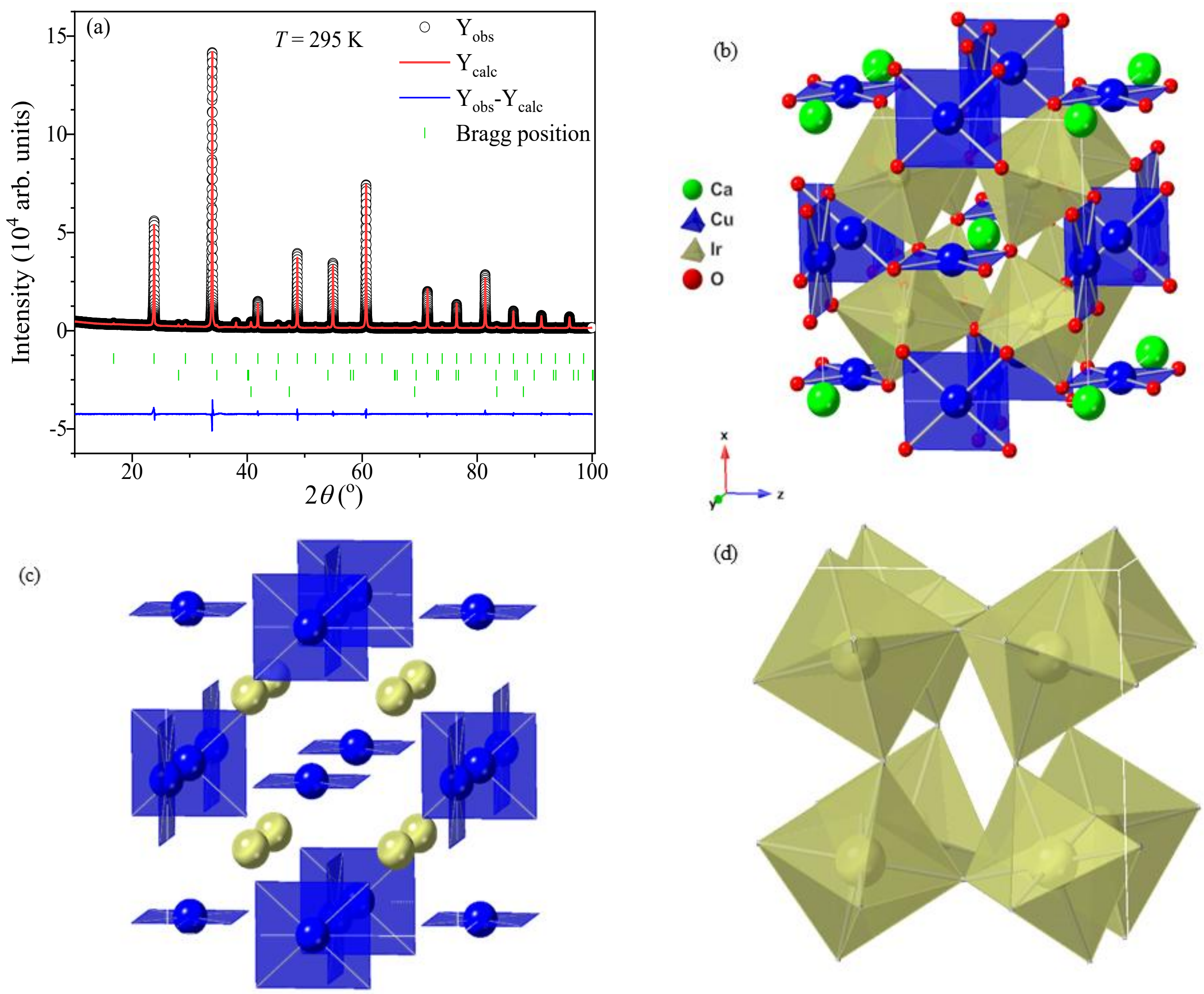

**Fig 1.** (a) Rietveld-refined powder X-ray diffraction pattern of $CaCu_3Ir_4O_{12}$ at room temperature. The observed and calculated data are, respectively, represented by open black circles and red solid line, the difference between observed and calculated patterns by solid blue line. The top, middle, and bottom vertical green ticks, respectively, indicate the Bragg positions of $CaCu_3Ir_4O_{12}$, $IrO_2$, and Ir-metal phase. (b) Crystal structure of $CaCu_3Ir_4O_{12}$. (c) Cu planes with Ir atoms, and (d) corner-shared $IrO_6$ octahedral units.

## 3.2 Magnetic Properties

Temperature Dependence DC Magnetic Response

We first establish the strength of magnetic interactions using DC susceptibility measurements. Given that, $IrO_2$ is Pauli paramagnetic with only weak *T*-dependence of susceptibility and very small low-*T* value (~ 5 – 6 X $10^{-4}$ emu/mol at 2 K) while metallic Ir is a weak Pauli paramagnet with nearly temperature-independent susceptibility and even smaller low-*T* susceptibility magnitude (~ 2 X $10^{-4}$ emu/mol) [28], and that both these phases occur in very small quantities, neither of these two phases should have any influence to the ground state magnetism of $CaCu_3Ir_4O_{12}$. Therefore, the magnetic ground state responses, discussed below, solely represent the intrinsic nature of bulk magnetism of $CaCu_3Ir_4O_{12}$. Figs. 2(a) and S1 of supplementary materials show the temperature dependence of bulk DC magnetic susceptibility [$\chi(T)$] for $CaCu_3Ir_4O_{12}$ measured under magnetic fields from 0.5 kOe to 50 kOe. Across all fields, the ZFC and FC curves coincide across the full temperature range, with no divergence or anomaly down to 2 K, indicating the absence of spin freezing or long-range magnetic order and consistent with a dynamically correlated paramagnetic ground state [2,29]. The overall $\chi$ ($T$) profile decreases smoothly on cooling, without any anomaly associated with magnetic ordering [30]. Because the Curie–Weiss (CW) fitting parameters in correlated disordered magnetic materials are sensitive to both the chosen temperature range and the applied magnetic field, the CW analysis was carried out using field-cooled susceptibility data measured at high field within the field-independent high-temperature regime, following the modified Curie–Weiss expression Eqn. 1 [31].

$$\chi(T) = \chi^0 + \frac{C}{(T-\Theta_W)} \quad (1)$$

where $\chi^0$ is the temperature-independent contribution, $C$ is the Curie constant, and $\Theta_W$ is the Weiss temperature. Figs. 2 (a) and S1 also reveal the CW fitting results, giving $\Theta_W$ ~ -200 K (ranging between -185 and -225 K depending on applied field), and an effective magnetic moment, $\mu_{eff} = \sqrt{8C} \sim 4$ $\mu_B$/f.u. As shown in the inset of Fig. 2(a), $(\chi - \chi_0)^{-1}$ varies approximately linearly with temperature in the high-temperature regime. Clearly, below ~100 K, deviations from the CW behaviour become evident and persist at all applied fields, suggesting an intrinsic origin rather than a field-induced effect [32]. The large negative $\Theta_W$ (~ -200 K) indicates predominant antiferromagnetic (AFM) exchange interactions in this material, although this Weiss temperature is considerably smaller than the reported $\Theta_W \approx$ -397.9 K of the La analogue $LaCu_3Ir_4O_{12}$ system, where $Cu^{2+}$ moments order antiferromagnetically at a low temperature [23]. This contrast in magnitude may indicate that additional competing effects influence the low-temperature magnetic state in $CaCu_3Ir_4O_{12}$ and hinder the conventional long-range order. In addition, the estimated effective paramagnetic moment, $\mu_{eff}$ ~ 4 $\mu_B$/f.u., is larger than the theoretical spin-only value (~3 $\mu_B$) expected from three $Cu^{2+}$ ions per formula unit of $CaCu_3Ir_4O_{12}$ [33]. This enhancement suggests a departure from a picture based solely on strictly localized $Cu^{2+}$ moments. In contrast, the reported effective moments, $\mu_{eff} \approx 3.0 – 3.3$ $\mu_B$/f.u., of the wider $CaCu_3B_4O_{12}$ series ($B$ = Ge, Ti and Sn) are broadly consistent with the expected contribution from three $Cu^{2+}$ ions per formula unit, whereas the $CaCu_3Ru_4O_{12}$ exhibits a much smaller effective moment ($\mu_{eff}$ = 1.5 $\mu_B$/f.u.) due to strong Cu–Ru hybridisation and partial screening of the Cu spins [25,34,12]. Considering the fact that Ir 5d states in $CaCu_3Ir_4O_{12}$ have been reported from DFT studies to carry only small magnetic moments (~0.45 $\mu_B$), the observed enhancement of effective moment in our $CaCu_3Ir_4O_{12}$ system is more likely to reflect additional orbital contributions and Cu–Ir hybridization involving the spin–orbit-coupled Ir 5d states [9]. Therefore, although $Ir^{4+}$ in $CaCu_3Ir_4O_{12}$ is expected to be largely itinerant with negligible free-ion local moment, additional orbital and SOC contributions from the hybridized Ir 5$d$-O 2p network enhance the effective moment [4,21,23]. Additionally, as seen in Figs. 2a and S1 of SM, the field-independent broad feature of the $\chi(T)$ data centered at around 100 K, together with the departure of the CW law below ~100 K, might be an indication of magnetic coupling of $Cu^{2+}$ moments to the Ir-5d orbitals. Taken together, these results suggest a delicate interplay between SOC, electron correlation, 3$d$-5$d$ orbital hybridization, and competing exchange interactions towards the ground state magnetism of the $CaCu_3Ir_4O_{12}$ sample.

To examine the low temperature magnetic response, Fig. 2(b) shows the field-cooled $\chi(T)$ data in a reduced temperature window on a log-linear scale. Instead of following a Curie-tail, the low-temperature (< 15 K) susceptibility follows a sub-Curie power-law form, $\chi$ ~ $T^{-\alpha_s}$, with two distinct regimes: 1) In the range 6–15 K, the exponent $\alpha_s$ remains nearly constant at 0.25–0.26 under $H \leq 15$ kOe, but decreases to 0.22 and 0.19 at 30 and 50 kOe, respectively, and 2) Below 4 K, the field dependence becomes much stronger, with $\alpha_s$ decreasing from 0.25 at 15 kOe to 0.128 at 30 kOe and 0.06 at 50 kOe. These results show that the low-temperature bulk DC susceptibility is progressively suppressed with increasing field, consistent with field-induced polarisation of fluctuating local moments governed by low-energy spin excitations rather than the development of static magnetic order [4,29]. Moreover, as $T \rightarrow 0$, the $\chi(T)$ under the highest applied field of 50 kOe tends to saturate towards a finite value rather than decreasing to zero, which is reminiscent of the low-energy spin response reported in QSL-like systems [35].

Further, the field-dependent isothermal magnetizations (Fig. 2c) reveal no detectable hysteresis or coercivity or remanent magnetization up to ±60 kOe down to the lowest measured 2 K, ruling out any bulk ferromagnetic component (or field-trained moment) in this compound. This is further corroborated by the negative interceptions of the 2 and 5 K Arrott plots ($M^2$ versus $H/M$) onto $M^2$ - axis (Fig. S2), which is in agreement with the absence of spontaneous magnetization vis-à-vis ferromagnetic components [36,37]. Notably, the nonlinear *M*-*H* behaviour becomes progressively stronger on cooling, without a sharp ordering signature, suggesting the response being dominated by field-polarized paramagnetic moments (or very weak short-range correlations) rather than long-range order. Consistently, the 2 K $M-H$ isotherm also follows a power-law form, $M \sim H^{1-\alpha_m}$, with $\alpha_m = 0.12$ for $H \leqslant 15$ kOe and 0.23 for $H > 15$ kOe, as shown in the inset of Fig. 2(c) rather than exhibiting a Brillouin-function-like paramagnetic dependence.

The consistency between the low-temperature power-law behaviours of $\chi(T)$ and $M(H)$, together with the absence of hysteresis or spontaneous magnetization, further argues against extrinsic paramagnetic centres. Instead, these results are consistent with a frustrated quantum magnetic ground state with abundant low-energy spin excitations, as widely discussed for systems featuring quantum spin liquid or random-singlet, or valence-bond-like states [38–40].

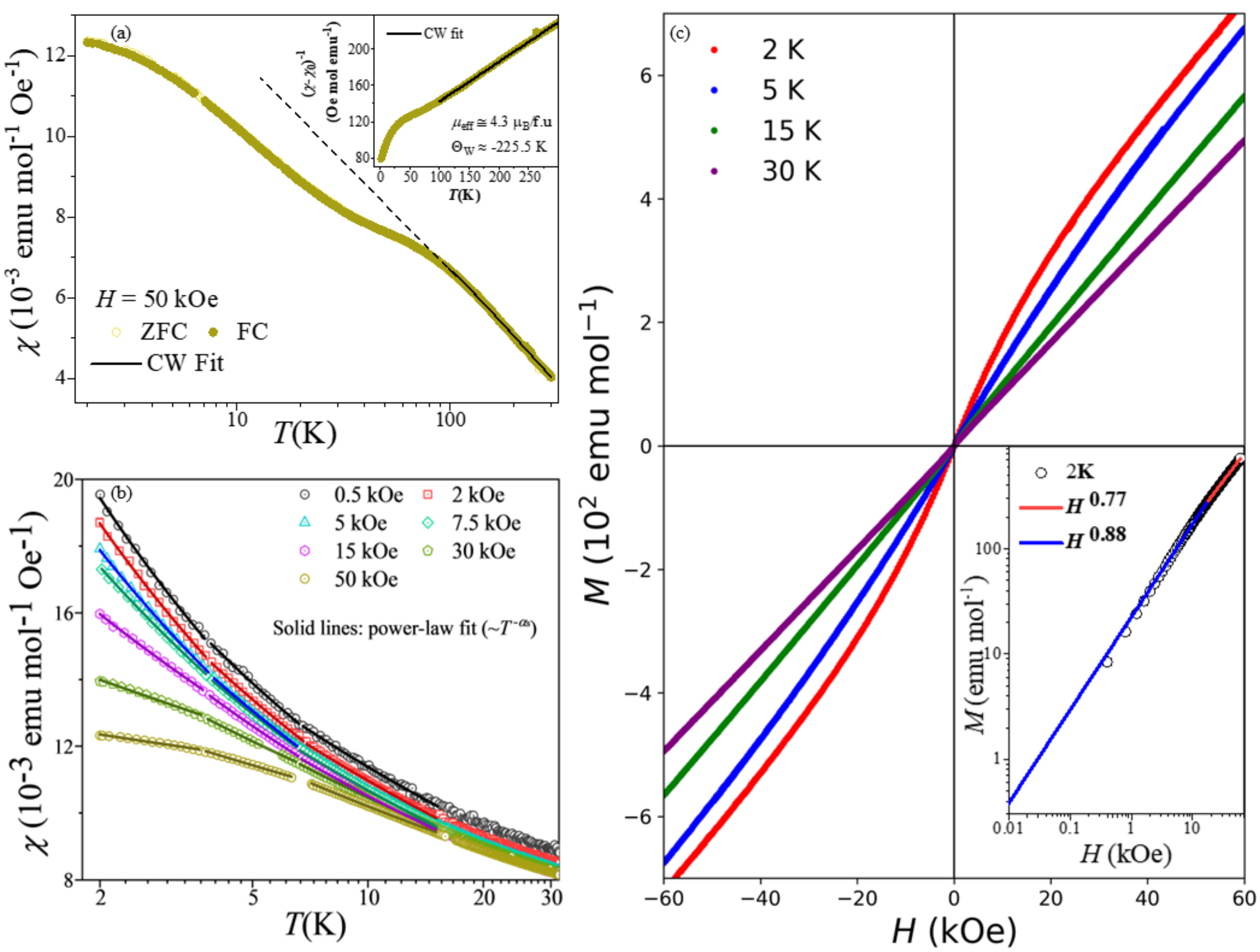

**Fig 2.** (a) Temperature dependence of the DC magnetic susceptibility, $\chi(T)$, measured at 50 kOe under zero-field-cooled (ZFC) and field-cooled (FC) conditions. The solid black line represents the Curie–Weiss fit. Inset: Respective inverse susceptibility ($[\chi - \chi^0]^{-1}$) versus temperature plot together with linear CW fitting.(b) Low-temperature FC susceptibility measured under applied magnetic fields from 0.5 to 50 kOe. Open symbols represent experimental data, while solid lines denote power-law fits performed over field-dependent temperature ranges selected for each dataset. where the fitted regions exhibit approximately linear behaviour. (c) Field-dependent isothermal magnetization $M(H)$ curves measured at some selected temperatures, showing reversible behaviour without hysteresis. Inset: Power-law scaling of the 2 K $M(H)$ isotherm shown on log–log scale, showing two power-law regimes.

Low-Temperature AC Magnetic Response

The AC susceptibility provides an additional probe of slow or frozen magnetic dynamics across a broad temperature window. Figs 3 (a-b) summarise $\chi'_{\mathrm{ac}}(T)$ measurements spanning 50 mK–300 K under multiple different frequencies. Throughout the entire temperature range, $\chi'_{\mathrm{ac}}(T)$ varies smoothly and monotonically, without any cusp-like anomalies. The collapse of $\chi'_{\mathrm{ac}}(T)$ curves of different frequencies onto a single, frequency-independent trajectory further confirms the absence of spin-frozen ground state or glassy spin dynamics. [41,42] Such behaviour contrasts sharply with canonical spin-glass or cluster-glass systems, which exhibit strong frequency dispersion governed by Vogel–Fulcher or critical slowing-down scaling. [2,43] Clearly, despite having sufficiently strong AFM exchange between the sizeable local moments, the $CaCu_3Ir_4O_{12}$ material shows no sign of magnetic ordering phase transition down to 0.05 K, giving rise to very high magnetic frustration, in terms of frustration index, $f = \frac{|\Theta_W|}{T_{lowest}} \approx 4000 >> 1$, in this system. Thus, together with the bulk DC susceptibility and magnetization results, these observations are consistent with $CaCu_3Ir_4O_{12}$ being a promising QSL candidate [35,44].

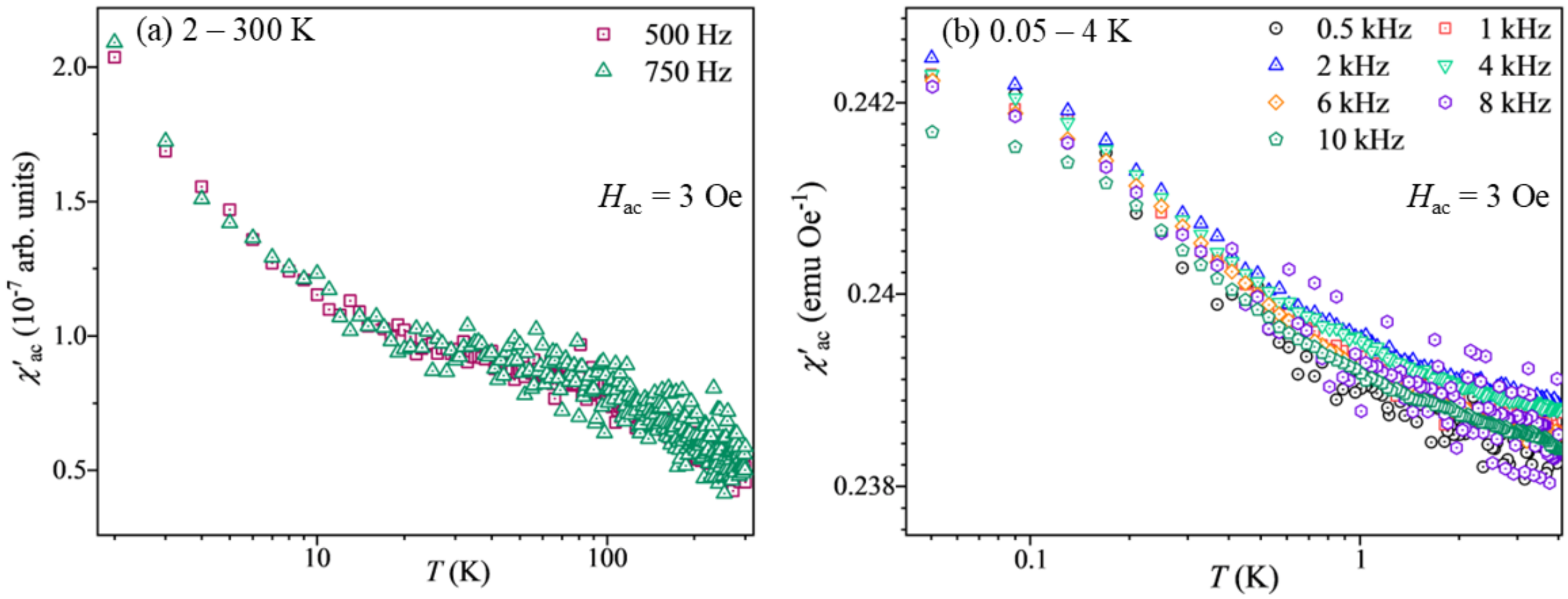

**Fig 3**. (a) Temperature dependence of real part of the AC magnetic susceptibility data measured between 2 and 300 K, and (b) 0.05 and 4 K with an excitation field of $H_{\mathrm{ac}} = 3$ Oe in selected frequencies.

## 3.3 Heat Capacity

To further investigate the magnetic ground state and low-energy spin-excitations, the specific heat was measured as a function of temperature under zero and various applied fields over the 2 - 150 K and 0.05 - 4 K temperature ranges, as shown in Figs. 4 (a) and

(b) respectively. No sharp $\lambda$-like anomaly is observed down to the lowest measured temperature, further supporting the absence of thermodynamic phase transition in $CaCu_3Ir_4O_{12}$. The specific-heat data plotted as $C_p/T$ [inset of Fig. 4a] exhibit a broad, field-independent maximum centred at around 85 K, in agreement with the previous work by Cheng et al [13] and also with the reported behaviours of the heavy-fermion Ru-analogue $CaCu_3Ru_4O_{12}$ [10]. Notably, this broad maximum appears in the same temperature region where the $\chi(T)$ data begin to deviate from the CW law behaviour on cooling, indicating the development of short-range magnetic correlations without the onset of long-range order at intermediate temperatures. The absence of any magnetic phase transition, together with nearly overlapping $C_p\,(T)$ curves under all different fields, is consistent with the thermodynamic signatures observed in the reported several quantum-disordered iridates [2,4,21,23]. In addition, below 10 K, the weak upturn in the $C_p/T$ data (Figure 4a, inset) may arise from the low-energy spin-fluctuations contribution or the Kondo coupling of Cu 3*d* moments with Ir 5*d* orbitals [12,13].

We now turn to the low-temperature (0.05-4 K) specific heat behavior. After subtracting the phonon contribution, taken as $\beta T^3$ with $\beta$ ~ 5 X $10^{-5}$ J $mol^{-1}$ $K^{-4}$, the resulting magnetic specific heat, $C_{mag}$, measured under both zero and applied fields is displayed in Fig. 4 (b). As seen in both the main panel and the inset of Fig. 4b, there is no field-dependent anomaly or broad feature in $C_{mag}/T$ data throughout the measured temperature range up to the highest applied field of 90 kOe, arguing against a “two-level Schottky anomaly” effect in this sample. Consistently, failure to model the difference specific heat, $\Delta C_p(H,T) = C_p(H,T) - C_p(0,T)$, using a two-level Schottky model further supports the conclusion that paramagnetic impurities do not play a significant role in the ground-state magnetism of $CaCu_3Ir_4O_{12}$. Moreover, $C_{mag}/T$ shows a low-temperature upturn below about 0.5 K [inset of Fig. 4(b)] in both zero field and applied fields, with the magnitude of the upturn becoming progressively enhanced at higher fields. Such behaviour is commonly discussed in QSL candidates in connection with enhanced low-energy spin excitations [45,46], although its microscopic origin requires further in-depth investigation. Also, the release of magnetic entropy, obtained by integrating the $C_{mag}/T$ data with temperature [Fig. 4(c)], reaches only ~10.5–12% of the expected full value of $3R\ln 2 \approx$ 17.3 J mol$^{-1}$ K$^{-2}$ for three $Cu^{2+}$ ($3d^9$, $S$ = ½) spins per formula unit and within the consideration of largely itinerant Ir 5$d$ states. This suggests that most of the magnetic entropy (~90%) remains unreleased down to the lowest measured temperature, consistent with persistent spin fluctuations and low-energy spin excitations in the magnetic ground state [44,45].

Finally, to gain insight into the nature of low-energy spin excitations of this material, the 0.05 – 4 K $C_{mag}$-$T$ data in Fig. 4(b) were analysed using phenomenological expression:

$$C_{mag} = \gamma T + \alpha T^2 + \delta T^{-2} + D * T(-logT) \quad (2)$$

Here, $\gamma$ is the Sommerfeld coefficient, typically observed for metals, $\alpha$ is quadratic $T$ coefficient. and the third term, $\delta T^{-2}$ accounts for the nuclear Schottky contribution, where the coefficient '$\delta$' varies approximately as square of the applied field, $\delta \sim H^2$ [47], as illustrated in Fig. 4(e). The last term describes the negative logarithmic temperature dependence of $C_{mag}/T$, which together with only negligible field-dependence of the associated coefficient, '$D$' [Fig. 4f], might be an indication of the emergence of quantum criticality in this system [48,49]. Importantly, a satisfactory fit to the $C_{mag}$-$T$ data [Figs. 4 (b) and (d)] requires combined linear- and quadratic-temperature-dependent terms, *i.e.*, $C_{mag} \sim \gamma T + \alpha T^2$. The linear temperature-dependence of $C_{mag}$, together with the astonishingly large Sommerfeld coefficient [$\gamma$ ~ 488 mJ mol$^{-1}$ K$^{-2}$ for $H$ = 0 and $\gamma_{avg}$ ~ 518.5 mJ mol$^{-1}$ K$^{-2}$ for $H$ ≥ 10 kOe, left y-axis of Fig. 4(d)], is likely in agreement with heavy fermion behavior associated with Kondo hybridization of Cu-3d localized moments and Ir-5d itinerant electrons [13]. On the other hand, this $T$-linear term is quite common in case of the gapless QSL ground state and has been attributed to the gapless low-energy spin-excitations from a metal-like spinon fermi surface of the QSL ground state [50–52]. As $CaCu_3Ir_4O_{12}$ is a metallic heavy fermion, hence, both spin and charge degrees of freedom will contribute to the low-temperature heat capacity, naturally leading to the large $T$-linear contribution. The quadratic $T$-dependency of $C_{mag}$ at very low temperatures, along with the finite $T^2$ term [$\alpha_{avg} \approx$ 14.5 mJ mol$^{-1}$ K$^{-3}$ for $H \leq$ 50 kOe and ~ 32 mJ mol$^{-1}$ K$^{-3}$ at $H \geqslant$ 70 kOe, shown by right y-axis of Fig. 4d], could be the consequence of novel gapless Dirac spinon excitations with linear dispersion, similar to the recently reported Dirac QSL materials $Sr_3CuSb_2O_9$ [53] and $YbZn_2GaO_5$ [54]. Moreover, this low-temperature $T^2$-dependence of magnetic specific heat is often described as a signature of gapless spin-excitation spectrum of frustrated quantum magnets [35,38]. To further assess whether the observed low-energy behaviour could arise from a randomness-driven random-singlet-state (RSS), we also tested the universal scaling relations among $\chi(T)$, $M(H)$, and $C_{mag}/T$ (see Section S3 in SM). However, no universal scaling or data collapse is observed for a common scaling exponent, making an RSS interpretation unlikely.

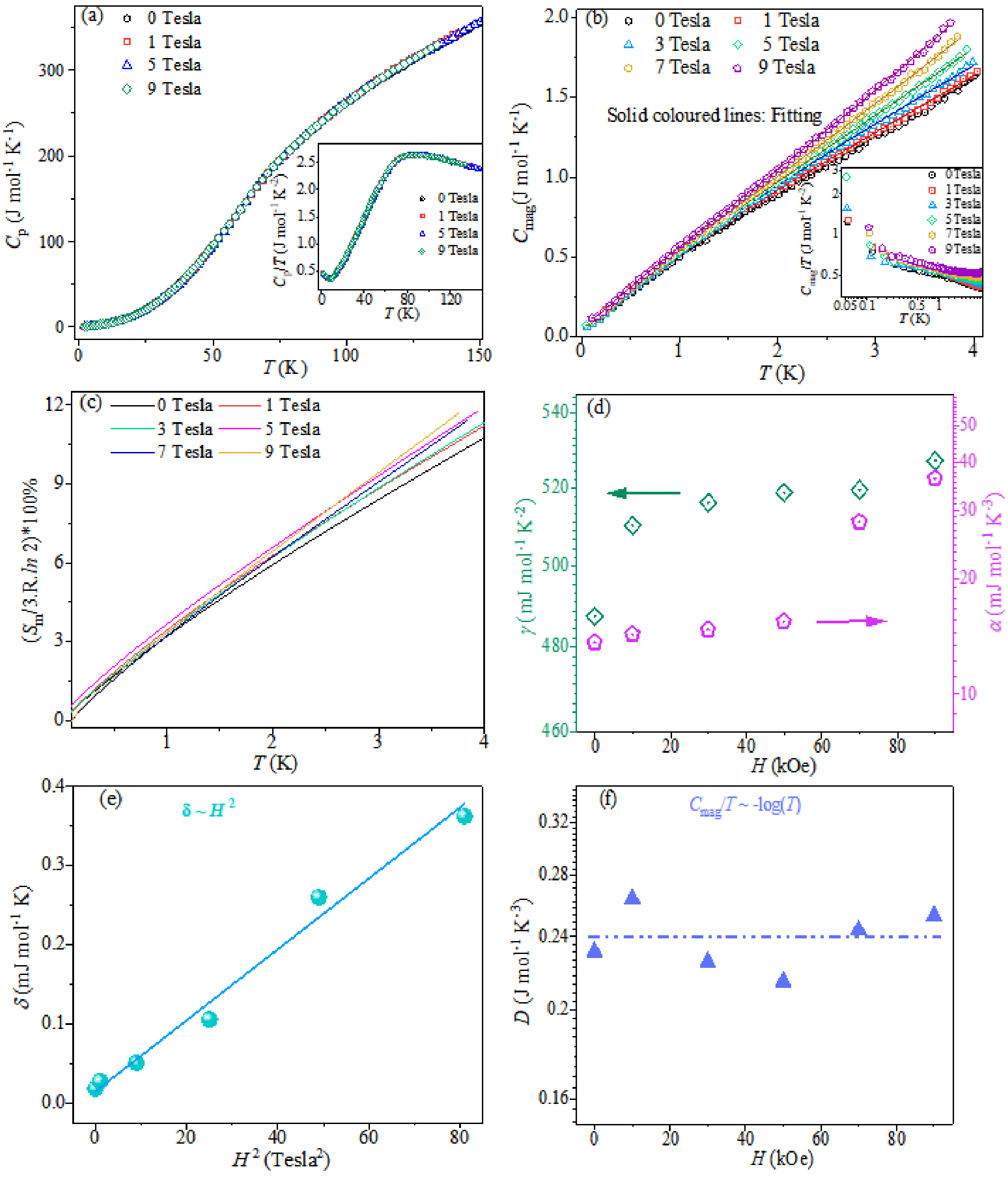

**Fig 4.** (a) Temperature dependence of total specific heat $C_p$ measured from 2–150 K in zero and applied different magnetic fields, Inset: Respective $T$-divided $C_p$ data plots against temperature. (b) Magnetic specific heat ($C_{mag}$) in the 0.05 – 4 K range under zero and applied fields. Solid coloured lines represent respective theoretical fits using Eqn. 2. (c) Magnetic entropy ($S_m$) in the 0.05 – 4 K range in both zero and applied fields. (d) Field dependence of $\gamma$ (left y-axis) and $\alpha$ (right y-axis), (e) Field-square dependence of nuclear Schottky term ($\delta$), and (f) Field dependence of the coefficient $D$ associated with the $T(-logT)$.

## 3.4 Muon Spin Relaxation (μSR)

Bulk magnetic and thermodynamic measurements show no evidence of long-range magnetic order or spin freezing down to 0.05 K, despite significant antiferromagnetic interactions. However, bulk probes alone cannot definitively establish the magnetic ground state in weak or disordered systems. At this point, Muon spin rotation/relaxation, μSR, as a highly sensitive local probe capable of detecting extremely small internal fields (of the order of 0.1 Oe) arising from any weak magnetic order or spin freezing, is therefore employed to determine the magnetic ground state and low-energy spin dynamics of $CaCu_3Ir_4O_{12}$ [21,44,45,55].

Fig. 5 (a) illustrates the time evolution of the zero-field (ZF)-μSR asymmetry spectra along with fittings for some selected temperatures between 0.04 and 4 K. Clearly, the spectral line-shape undergoes a gradual changeover from a Gaussian-like to a Lorentzian-like relaxation form with decreasing temperature. Further, there is neither coherent spontaneous oscillations nor 2/3$^{rd}$ initial muon asymmetry drop (expected for larger magnetic moment ordering) followed by 1/3$^{rd}$ muon-polarization-recovery down to the lowest measured 0.04 K, thus ruling out static magnetic ordering (neither a long-range magnetically ordered state nor any partial spin-freezing) down to 0.04 K in $CaCu_3Ir_4O_{12}$.

All the ZF-μSR asymmetry curves have been satisfactorily fit using a single stretched exponential relaxation function as

$$A(t) = A_{\mathrm{rel}} \exp\left[-(\lambda t)^{\beta}\right] + A_{bkg} \tag{3}$$

Here, $A_{bkg}$ is the temperature-independent background arising from the muons stopping at the Ag-sample holder. $A_{rel}$, λ, and β are the muon asymmetry amplitude, relaxation rate, and stretched exponent, respectively, corresponding to the relaxation of Cu-electronic moment fluctuations. Considering itinerant character of the Ir magnetic moments and that μSR is particularly sensitive to the localized, slowly fluctuating moments over the itinerant and paramagnetic fluctuating electronic moments, the observed time-evolution of muon-spin relaxation as a function of temperature or longitudinal field in our $CaCu_3Ir_4O_{12}$ material is therefore wholly attributed to the fluctuations of Cu-electronic moments. $A_{bkg}$ and $A_{rel}$ were determined to be ~ 10.8 and 11.8, respectively, after fitting the base-T (0.04 K) μSR asymmetry curve and then kept fixed at the respective values throughout the remaining temperatures data analysis. The corresponding temperature evolution of the muon-spin relaxation rate, $\lambda$ and the associated stretching exponent, $\beta$, extracted from the fits to the ZF-muon asymmetry spectra using Eqn. 3, are shown by left and right Y-axes, respectively, of Fig. 5 (b).

Upon cooling from 4 K to 0.15 K, the temperature dependence of the relaxation rate, $\lambda$, gradually increases with decreasing temperature (Figure 5b), indicating a slowing down of Cu-spin fluctuations through the development of magnetic correlations, as commonly observed in reported QSL materials [34,52,53]. However, such an obvious slowing down of the spin dynamics couldn't result in the static magnetic ordering till down to the lowest measured temperature of 0.04 K, as reflected from the missing of diverging $\lambda$-$T$ variation or a peak in the $\lambda(T)$ curve even at the lowest $T$ [see left y-axis of Fig. 5 (b)]. Notably, upon temperature lowering below 0.15 K, the relaxation rate, $\lambda$, levels off at ~ 0.156 $\mu s^{-1}$ and maintains a nearly temperature-independent flat plateau-like behaviour in the $T$-window of 0.04 – 0.15 K. This endorses persistence of strong quantum spin fluctuations in this material down to at least 0.04 K, in agreement with a dynamic QSL ground state. [47,53,56]

The stretched exponent, $\beta$, gradually decreases [right y-axis of Fig. 5 (b)] on cooling from ~1.35 at 4 K, and finally achieves an almost constant value of $\beta$ ~ 0.82 – 0.83 between 0.15 and 0.04 K. Such a reduced $\beta$ from unity ($\beta$ = 1 corresponds to homogeneous magnetism) could point to the emergence of inhomogeneous magnetic environment in this material at low-$T$ [57,58], which might be due to the pertinent short-range magnetic correlations and also the Kondo coupling of Cu-3d moments with Ir-5d bands, disrupting the overall homogeneous magnetic environment at this low-$T$ and therefore, affecting the $\beta$-value. Notably, the low-$T$ saturated $\beta$-value (~ 0.82 – 0.83) is well above the $\beta$ = 1/3 of a typical spin-glass system. [59,60], thus rejecting further any spin-freezing in the magnetic ground state of $CaCu_3Ir_4O_{12}$. Again, the $\beta$ > 1 in the 0.5 – 4 K $T$-range may arise from a distribution of the muon stopping sites close to Ir ions, as discussed in case of the dimer iridate $Ba_3InIr_2O_9$. [58] Moreover interestingly, the temperature (~ 0.15 K), at which both $\lambda$ and $\beta$ start to level-off upon cooling, is same as those below which the real part of the AC magnetic susceptibility $\chi'_{ac}$(T) approaches nearly saturation [see Fig. 3c] and the $T$-divided magnetic specific heat, $C_{mag}/T$, reveals an upturn. This possibly attributes the characteristic temperature of 0.15 K to the onset of magnetic fluctuations in this material, and a strongly quantum-entangled spin-liquid-like phase emerging below this temperature.

Next, in order to further check if the spin-spin correlations are truly dynamic or static and also to identify whether the origin of the 0.04 – 0.15 K flat plateau of the ZF-muon spin relaxation rate is static or dynamic, the muon decoupling experiments were conducted at the base-$T$ of 0.1 K in various applied longitudinal fields (LFs) up to 4 kOe, and the representative LF-μSR asymmetry curves are shown in Fig. 5 (c) for some selected LFs along with the fittings. The same Eqn. 3 was used to analyse all the collected LF-μSR spectra. $A_{bkg}$ is kept fixed at the same ZF-value of 10.8 and $A_{rel}$ remains nearly independent (~ 11.6 – 11.9) in applied LFs till to 1.5 kOe. While beyond 1.5 kOe LF, the observed gradual vertical upward shift of the muon asymmetry spectra as a function of the increasing LF is not at all related to any sample's intrinsic effect, rather it is solely due to the instrument effect in higher LFs, common in ISIS Muon facility. In order to take into account this vertical upward shift in the initial muon asymmetry at higher LFs, we let $A_{bkg}$ free to be varied with LF at $H_{LF}$>1.5 kOe while keeping the $A_{rel}$ unchanged between 11.6 and 11.9 throughout the measured LF range. As presented in Figs. 5 (c) and (d), clearly, there is gradual decoupling of the muon spins from the effect of internal field with increasing LF, surprisingly however, complete suppression of muon depolarization is still missing even at the highest applied 4 kOe LF, and perceptibly finite muon-spin relaxation is present at this maximum applied LF along with the nearly unchanged relaxation rate, $\lambda$, at $H_{LF} \geq 3$ kOe. Furthermore, if the $T$-independent plateau of ZF-muon depolarization rate ($\lambda$) between 0.04 and 0.15 K arises from any static internal field of width $\Delta H$, then the size of that static local field would be determined as: $\Delta H = \lambda_{ZF}/\gamma_{\mu} \approx 1.8$ Oe, where $\gamma_{\mu} = 2\pi \times 135.5$ MHz/Tesla is the muon's gyromagnetic ratio and $\lambda_{ZF}$ is the ZF-muon spin relaxation rate at the lowest measured 0.04 K. Now, complete decoupling of the muon spins from the effect of static internal field should occur in applied external LFs of (5 – 10) $\times \Delta H$. [38,52,61]

Yet astonishingly, no sign of fully quenched muon-depolarization is evident in the highest applied 4 kOe LF which is about 2200 times larger than $\Delta H \approx 1.8$ Oe, endorsing the dominance of fluctuating local internal fields rather than the static magnetism towards the low-$T$ (0.04 – 0.15 K) muon depolarization in this sample, as also reported in other QSL materials. [60] In addition, the stretched exponent, $\beta$, initially falls down from ~ 0.825 at zero-field to ~ 0.42 for $H_{LF}$ = 25 Oe, and then, $\beta$ ranges only between 0.4 and 0.5 throughout the entire applied LF regime with no such field-dependency [see inset to Fig. 5 (d)], aligning neither with a paramagnet ($\beta \approx 1$ across all fields) nor with a spin-frozen phase ($\beta \sim 0.3$). [60,62] Together, all these corroborate the persistence of strongly quantum-fluctuating Cu-local moments till down to 0.1 K at least, thus validating a truly dynamic QSL ground state in $CaCu_3Ir_4O_{12}$.

Now for an exponentially decaying spin-spin correlation function, $S(t) \propto \exp(-\nu t)$, the LF-dependence of the muon spin relaxation rate can be described by the Redfield formalism (See Text S4 in SM). Such a correlation leads to the Lorentzian spectral density $S(\omega)$. [63,64] As depicted in Fig. 5(d), the Redfield formula provides only a very poor description of the longitudinal-field dependence of the muon spin relaxation rate, $\lambda_{LF}(H_{LF})$, implying a non-exponentially decaying spin-spin correlation function in our $CaCu_3Ir_4O_{12}$ system. [63] On the other hand, for a power-law decaying spin-correlation function, $S(t) \propto t^{-(1-\gamma)}$, the spectral density is given by $S(\omega) \propto \omega^{-\gamma}$ and consequently, the LF-dependence of the muon-spin relaxation rate can be best represented as, $\lambda_{LF} \propto H_{LF}^{-\gamma}$ . [63,65] As illustrated in Fig. 5 (d), unlike the Redfield description, the 0.1 K $\lambda$ – $H_{LF}$ data nicely follows the power-law dependence as $\lambda \propto H_{LF}^{-0.74}$. This sublinear power-law behaviour points to the unconventional spin correlation and spectral density of forms $S(t) \propto t^{-0.26}$ and $S(\omega) \propto \omega^{-0.74}$, respectively, in our $CaCu_3Ir_4O_{12}$ material. Such an algebraically decaying spin-spin correlations have been reported in various QSL materials with spin-chain, pyrochlore, Kagome, and triangular lattices. [55,65,66]

Moreover, the collected 0.1 K LF-μSR spectra under applied longitudinal fields up to 1.5 kOe are well described by a scale-invariant time-LF scaling relation of the type, $A(t, H_{LF}) = A(t/\mu_0 H_{LF}^{\gamma})$, as presented in Fig. 5 (e). Clearly, a reasonably good scaling is achieved over more than 4 orders of magnitude of $t/\mu_0 H_{LF}^{\gamma}$ for $\gamma = 0.74$. Such a scale-invariant time-field scaling relation of the 0.05 K LF-μSR asymmetry spectra, together with the lack of universal scaling and data collapse of $\chi(T)$, $M(H)$, and $C_{mag}/T$ [see in Sec. S3 of SM], therefore corroborates a dynamically correlated true QSL ground state of $CaCu_3Ir_4O_{12}$ and argues against an RSS as observed in the recently reported frustrated quantum magnets [63,67,68]. This further reinforces the view that the fluctuation-dominated QSL-like state of $CaCu_3Ir_4O_{12}$ is not disorder/randomness-driven but, instead, arising from intrinsic electronic correlations. Such universal scaling is also a characteristic signature of the QSL candidates. Additionally, the 1 K and 3.5 K muon decoupling measurements were performed, and the corresponding LF-μSR asymmetry spectra were analysed using the same Eqn. (3). The results for some selected LFs are displayed in Fig. S4. As decorated in Figs. S4 (c) and (d), the LF-dependence of the muon-spin relaxation rate, $\lambda$, and the stretched exponent, $\beta$, obtained from both the 1 and 3.5 K LF-μSR data fittings, are plotted and compared with those of the lowest $T$ (0.1 K) muon decoupling results.

All the qualitative and quantitative findings of our in-depth ZF and LF-μSR investigations provide strong evidence for a QSL ground state in $CaCu_3Ir_4O_{12}$. At this point, we may propose that the long-range RKKY interaction, activated between the Cu-local moments via Ir-5d itinerant bands, brings alternating ferro and antiferromagnetic exchange couplings across the lattice spatial distances of this heavy-fermion material, which could

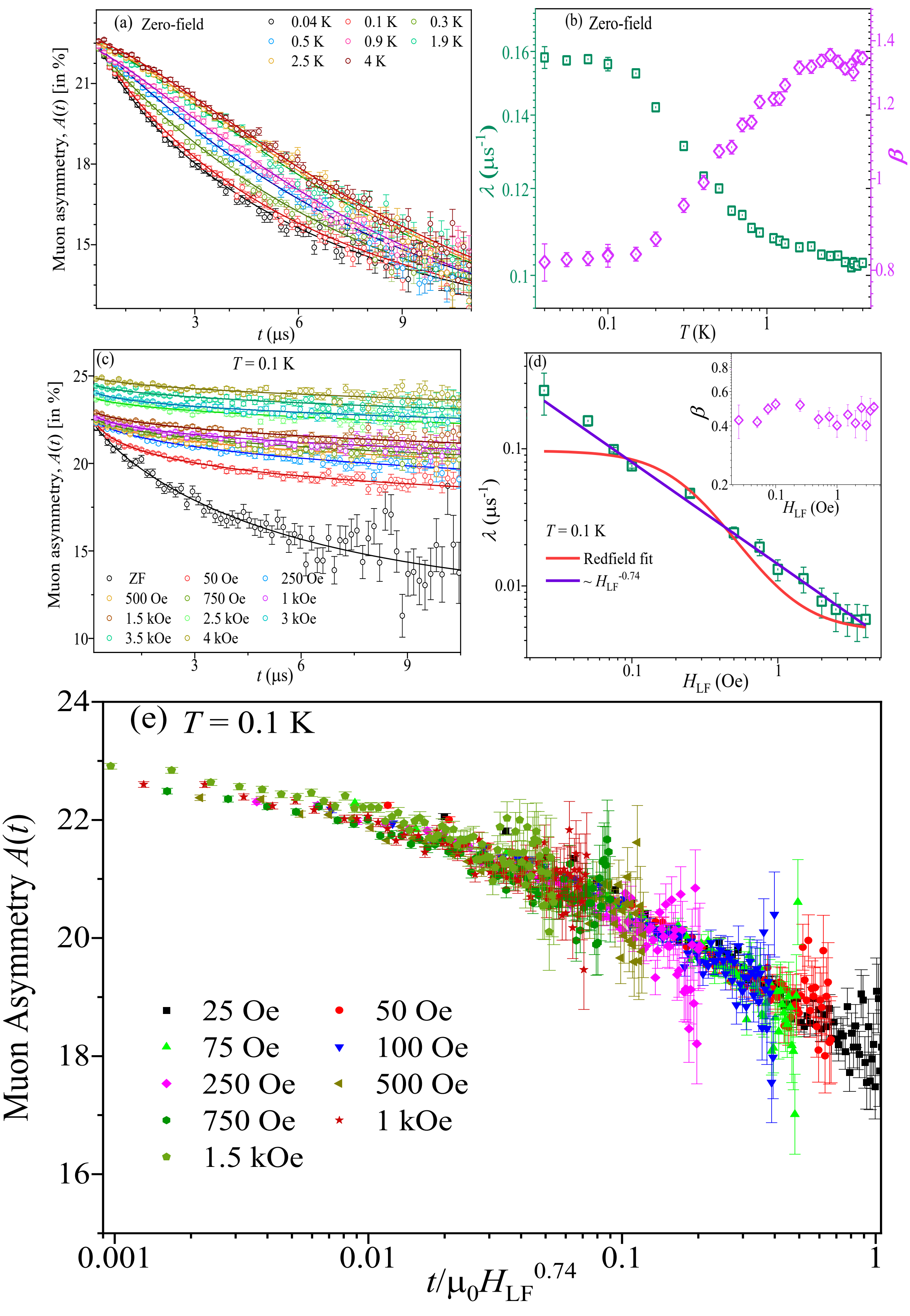
(a) Zero-field
0.04 K
0.1 K
0.3 K
0.5 K
0.9 K
1.9 K
2.5 K
4 K
Muon asymmetry, A(t) [in %]
t (μs)
(b) Zero-field
λ (μs⁻¹)
β
T (K)
(c)
T = 0.1 K
ZF
50 Oe
250 Oe
500 Oe
750 Oe
1 kOe
1.5 kOe
2.5 kOe
3 kOe
3.5 kOe
4 kOe
(d)
T = 0.1 K
Redfield fit
~ H_LF^-0.74
H_LF (Oe)
(e) T = 0.1 K
25 Oe
50 Oe
75 Oe
100 Oe
250 Oe
500 Oe
750 Oe
1 kOe
1.5 kOe
Muon Asymmetry A(t)
t/μ0H_LF^0.74

**Fig 5.** (a) Time evolution of the zero-field (ZF)-μSR asymmetry spectra (shaded coloured circles) at some selected temperatures, along with the fits (solid coloured lines); (b) Temperature dependence of the zero-field muon-spin relaxation rate, $\lambda$ (Left Y-axis with shaded olive squares) and the associated stretched exponent, $\beta$ (right Y-axis with shaded magenta kites) on a log-log scale; (c) Time evolution of the longitudinal-field (LF)-μSR asymmetry spectra at the lowest measured 0.1 K along with the fittings (solid coloured lines) for some selected LFs. (d) LF-dependence of the muon-spin relaxation rate, $\lambda$, along with the Redfield fitting (solid red line) and power-law dependence (solid blue line), shown on a log-log scale (main panel). Inset: LF-dependence of the stretched exponent $\beta$ on a log-log scale; (e) Time-LF scaling plot of the LF-μSR asymmetry curves at $T$ = 0.1 K.

be responsible for offering the competing exchange interactions and hence, establishing the magnetic frustration in this system. The exchange frustration together with low Cu-spin (S = ½), would therefore facilitate enhanced quantum fluctuations among the sizeable local moments, thereby favouring stabilization of a dynamic QSL ground state in this compound. Another perspective could also be posed here that the coexistence of $Cu^{2+}$ sublattice with an extended Ir–5d framework might possibly offer a delicate interplay between spin–orbit coupling and Cu–Ir hybridisation that could act as an important role in sustaining the dynamic spin correlations in this compound.

# 4. Conclusions

Our comprehensive bulk DC, AC magnetic susceptibility, specific heat, and local magnetic probe μSR characterizations provide coherent experimental picture that evidence missing of any long-range magnetic ordering or spin-freezing in $CaCu_3Ir_4O_{12}$ down to 0.04 K at least, despite having sufficiently strong antiferromagnetic interactions ($\Theta_W$ being in the range of ~-200 K depending on the applied field) between the sizable local moments. On top of it, our combined ZF and LF-μSR results confirm persistent spin dynamics down to at least 0.04 K, which together with universal scaling analysis certify the formation of a truly dynamic QSL ground state in this material. Further, the linear and quadratic temperature-dependence of magnetic specific heat is suggestive of low-lying gapless spinon excitation spectrum in the QSL ground state. Taken together, we hope our present work would offer a potential materials-avenue within the mixed 3d/5d heavy-fermion family of chemically ordered correlated 3D quadrupole perovskite oxides to explore in the QSL and frustrated quantum magnetism fields.

Further investigations using inelastic neutron scattering and resonance inelastic x-ray scattering will provide precise estimation of the magnetic exchange interaction strengths, magnetic anisotropy, spin-orbital entanglement, low-energy excitations, etc. in order to gain microscopic insights into the proposed magnetic frustration and consequently, the origin of the unconventional low-*T* fluctuating spin-liquid phase in this three-dimensional chemically ordered mixed 3d/5d magnetic oxide.

## 5. Acknowledgment

J. M., A. B., and D. T. A. thank EPSRC UK for the funding (Grant No. EP/W00562X/1). J.G.C. is supported by the National Key Research and Development Program of China (2023YFA1406100 and 2021YFA1400200). A. B. and D. T. A. would further like to thank the Royal Society of London for International Exchange funding between the UK and Japan, and Newton Advanced Fellowship funding between UK and China and the CAS for PIFI Fellowship. A.B. thanks the Lalit Narayan Mithila University, Darbhanga, India, for financial support. The authors acknowledge the Materials Characterization Lab (MCL) of ISIS facility, UK, for providing the experimental facilities. We thank the ISIS Facility for beam time on the EMU spectrometer RB2310680: Dr Devashi Adroja et al; (2024): Probing local spin dynamics in a novel d-electron heavy-fermion iridate $CaCu_3Ir_4O_{12}$ using muon-spin-relaxation/rotation (μSR), STFC ISIS Neutron and Muon Source, https://doi.org/10.5286/ISIS.E.RB2310680

# Supplementary Materials:

## Emergent Dynamic Magnetic Ground State in a Mixed 3d/5d Heavy Fermion System $CaCu_3Ir_4O_{12}$

**J. Ming[1, +, #], Abhisek Bandyopadhyay[1,2, *, #], G. B. G. Stenning[1], M. T. F. Telling[1], N. N. Wang[3,4], G. Wang[3,4], J.-G. Cheng[3,4] and D. T. Adroja[1,5, $]**

**[1]ISIS Neutron and Muon Source, STFC, Rutherford Appleton Laboratory, Chilton, Oxon OX11 0QX, United Kingdom**

**[2] Department of Physics, Ramashray Baleshwar College (A Constituent Unit of Lalit Narayan Mithila University, Darbhanga, India), Samastipur, Bihar 848114, India**

**[3]Beijing National Laboratory for Condensed Matter Physics, and Institute**

**of Physics, Chinese Academy of Sciences, Beijing 100190, China.**

**[4]School of Physical Sciences, University of Chinese Academy of Sciences, Beijing 100190, China**

**[5]Highly Correlated Matter Research Group, Physics Department, University of Johannesburg, Auckland Park 2006, South Africa**

+jingming1017@gmail.com

*abhisek.ban2011@gmail.com

$devashibhai.adroja@stfc.ac.uk

#J.M. and A.B. equally contribute to this work

**Date:6th May 2026**

**Summary:** This Supplementary document presents refined structural parameters from PXRD data refinement, along with the additional details of the DC magnetic susceptibility, isothermal magnetization, and longitudinal-field (LF)-$\mu$SR measurements for $CaCu_3Ir_4O_{12}$. Curie–Weiss fitting and Arrott-plot analysis are used to characterise the magnetic interactions, while field–temperature scaling of the thermodynamic quantities, together with comparative LF-$\mu$SR analyses, is used to further assess the nature of the magnetic ground state and spin dynamics.

**Table S1.** Crystal and refinement parameters of PXRD data for $CaCu_3Ir_4O_{12}$ at 295 K.

| Phases | Phase 1 (Major) | Phase 2 (Minor) | Phase 3 (Minor) |
|---|---|---|---|
| **Formulas** | $CaCu_3Ir_4O_{12}$ | $IrO_2$ | Ir |
| **Weight Fract (%)** | 96.67(14) | 1.28(5) | 2.06 (5) |
| **Space Group** | *I m* -3 | *P* $4_2$/*m n m* | *F m* -3 *m* |
| **Z** | 2 | 2 | 4 |
| **$D_{cal}$(g/cm$^3$)** | 9.488 | 11.681 | 22.559 |
| **Cell Parameters** | $a = b = c = 7.47142(4)$ Å<br>$\alpha = \beta = \gamma = 90^o$ | $a = b = 4.49597(46)$ Å,<br>$c = 3.15368(46)$ Å<br>$\alpha = \beta = \gamma = 90^o$ | $a = b = c = 3.83942(21)$ Å<br>$\alpha = \beta = \gamma = 90^o$ |
| **Volume (Å$^3$)** | 417.070(4) | 63.748(13) | 56.598(5) |
| **Bragg R-factor (%)** | 4.59 | 26.6 | 3.60 |
| **R-factors** | $R_p$ = 3.37 %, $R_{wp}$ = 5.08 %, $R_{exp}$ = 1.88 %, $\chi^2$ = 7.32 | | |
| **No. of variables** | 64 | | |
| **No. of profile points** | 19194 | | |

**Table S2.** Refined structural parameters for $CaCu_3Ir_4O_{12}$ at 295K.

| | Atoms | Site | *x* | *y* | *z* | Occ. | Biso (Å)$^2$ |
|---|---|---|---|---|---|---|---|
| **Phase-1 (Major) $CaCu_3Ir_4O_{12}$** | Ca | 2*a* | 0 | 0 | 0 | 1.0 | 0.022 |
| | Cu | 6*b* | 0 | ½ | ½ | 1.0 | 0.015 |
| | Ir | 8*c* | ¼ | ¼ | ¼ | 1.0 | 0.051 |
| | O | 24*g* | 0.30646 | 0.17305 | 0 | 1.0 | 0.026 |
| **Phase-2 (Minor) $IrO_2$** | Ir | 2*a* | 0 | 0 | 0 | 1.0 | 0.045 |
| | O | 4*f* | 0.19989 | 0.19989 | 0 | 1.0 | 0.029 |
| **Phase-3 (Minor) Ir** | Ir | 4*a* | 0 | 0 | 0 | 1.0 | 0.500 |

## S1. Curie–Weiss Analysis

The effective magnetic moment and Weiss temperature were obtained by fitting the high-temperature susceptibility to the modified Curie–Weiss law according to Eqn. S(1).[21]

$$\chi(T) = \chi_0 + \frac{C}{T - \Theta_W} \qquad \text{S (1)}$$

where $\chi(T)$ is the molar DC susceptibility, $\chi_0$ is the temperature-independent DC susceptibility term, *C* is the Curie constant, and $\Theta_W$ *is* the Weiss temperature. $\chi_0$, *C* and $\Theta_W$ were treated as free parameters and obtained from a least-squares fit of $(\chi - \chi_0)^{-1}$ versus *T*. The effective magnetic moment per formula unit, $\mu_{eff}$, is then calculated from the Curie constant via Eqn S(2).[22]

$$\mu_{eff} = 2.828\sqrt{c}\ (\mu_B / f.u.) \qquad \text{S (2)}$$

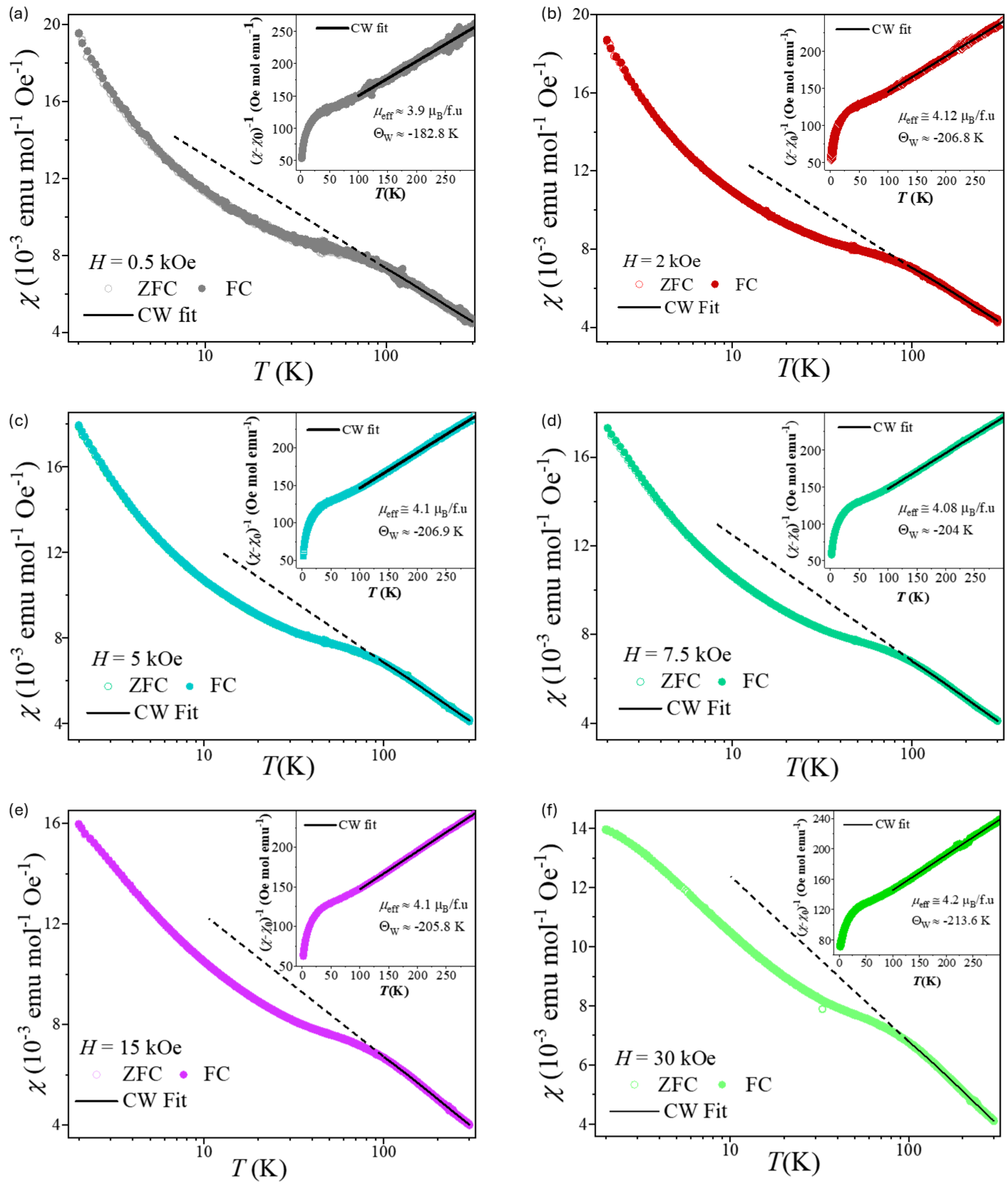

**Fig S1.** DC magnetic susceptibility, $\chi$(T), measured under a range of applied fields starting from 0.2 to 5 Tesla, showing both ZFC (open symbols) and FC (filled symbols) data. Solid coloured lines are the Curie–Weiss (CW) fits performed in the 100 – 300 K *T*-region (solid-coloured lines). Insets show the corresponding inverse susceptibility $(\chi - \chi_0)^{-1}$ *versus T* plots with linear CW fits used to extract $\mu_{eff}$ and $\theta_w$.

**S2. Arrott plot of the magnetization isotherms**

2 K and 5 K $M^2$ versus *H*/*M* Arrot plots were illustrated to explore the possibility of spontaneous magnetization behaviour.

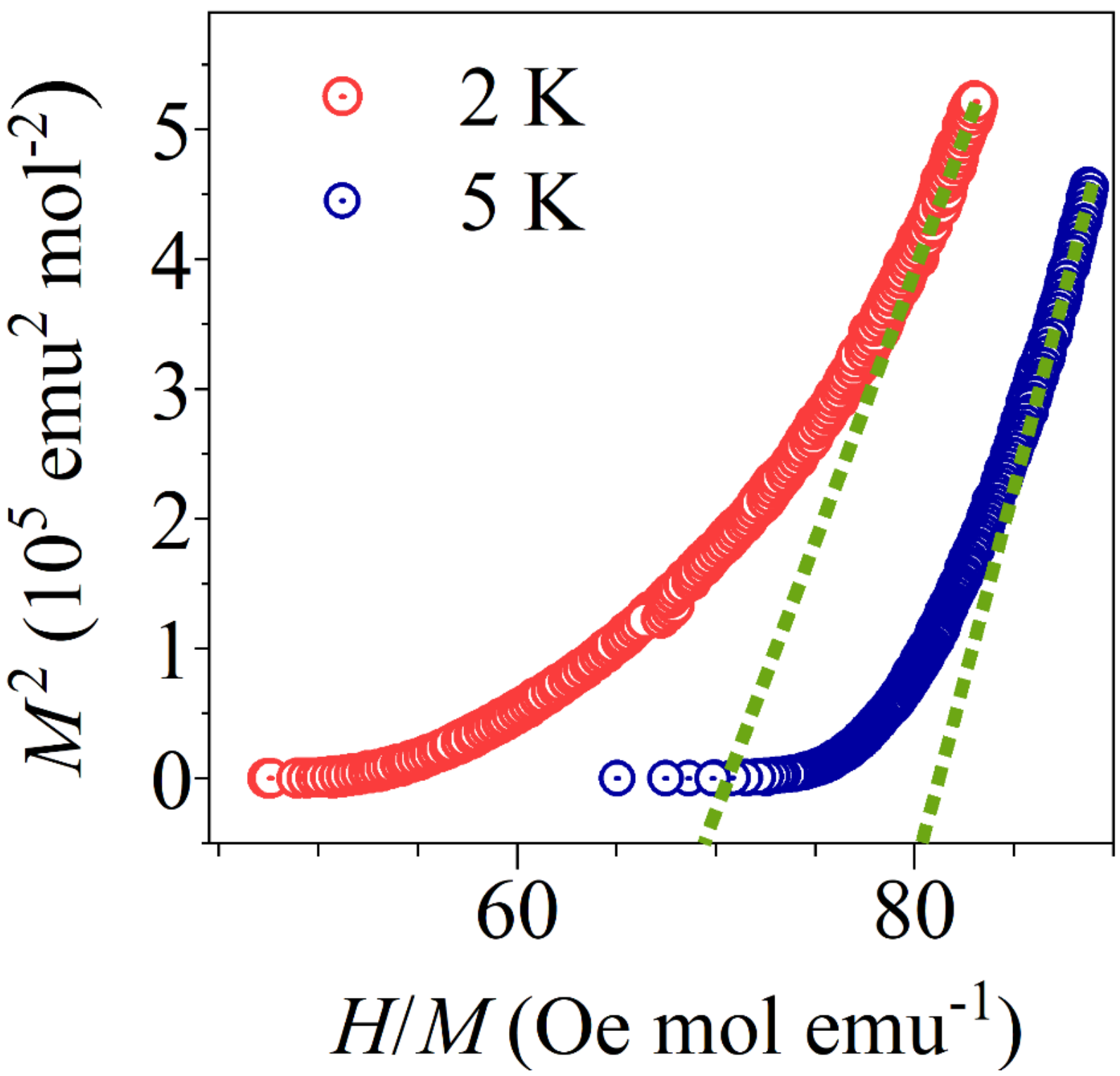

**Fig S2.** Arrott plots of the 2 and 5 K magnetization isotherms. $M^2$ is plotted as a function of $H/M$.

## S3. Field–temperature scaling analysis

As a randomness-driven random-singlet state (RSS) can share several low-temperature signatures with a quantum spin liquid (QSL), we tested the expected universal scaling behaviour among $\chi(T)$, $M(H)$, and $C_{mag}/T$. Experimental uncertainties were estimated to be approximately ±0.5% in temperature and ±0.01 Tesla in magnetic field. Repeated $C_p(T,H)$ measurements confirmed the reproducibility of the thermodynamic data.

The best fit results are depicted in Fig. S3, where both $H^{\alpha_s}\chi(T)$ and $H^{\alpha_h}C_{\mathrm{mag}}/T$ sets of data have been plotted separately against $T/H$ under different applied fields for the scaling exponent $\alpha_s$ = $\alpha_h$ = 0.3, as well as the $T^{(\alpha_m-1)}M(H)$ versus $H/T$ curves at different temperatures are presented on a same plot for the exponent $\alpha_m$ = $\alpha_s$ = 0.3. In contrast to the behaviour expected for an RSS, the $(H, T)$ dependence of $\chi(T)$, $M(H)$, and $C_{mag}/T$ data of $CaCu_3Ir_4O_{12}$ show neither universal scaling nor data collapse onto a single curve for a common scaling exponent.

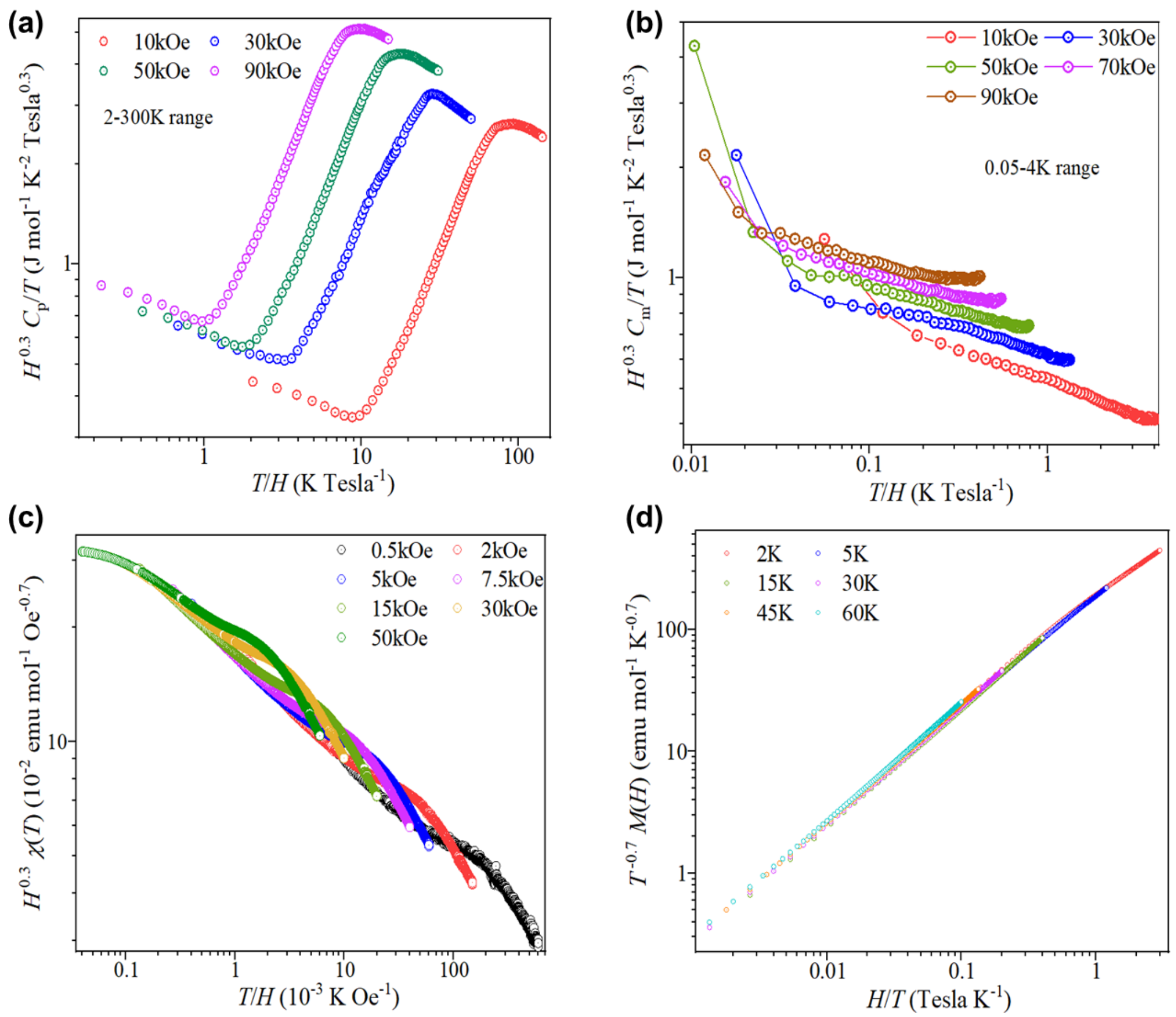

**Fig S3.** Field–temperature scaling of thermodynamic and magnetic responses in $CaCu_3Ir_4O_{12}$. Scaled specific heat plotted as ($H^{0.3}$ $C_p/T$) versus ($T/H$) over the 2–300 K range (a) and as ($H^{0.3}$ $C_{mag}/T$) versus ($T/H$) in the 0.05–4 K range (b). (c) Scaled DC susceptibility ($H^{0.3}$ $\chi(T)$) as a function of ($T/H$) for different applied magnetic fields, and (d) scaled magnetization isotherms plotted as ($T^{-0.7}M(H)$ versus ($H/T$)) at some different temperatures.

## S4. Analysis of the μSR data

Results from the 1 and 3.5 K muon decoupling experiments (*i.e.*, μSR measurements in applied various longitudinal-fields (LFs) at these fixed temperature) are presented in Fig. S4. The fitting parameters extracted from the respective LF-μSR data analyses are plotted against LF and compared with those of the 0.05 K results.

To investigate the nature of the spin dynamics, we analysed the longitudinal-field dependence of the μSR relaxation rate $\lambda(H_{LF})$. Here $S(t)$ denotes the spin–spin correlation function of the fluctuating local moments. An exponential form $S(t) \propto e^{-\nu t}$ corresponds to conventional stochastic spin fluctuations with a characteristic fluctuation rate $\nu$, leading to a Lorentzian spectral density and the Redfield field dependence of the μSR relaxation rate. In this case, the field dependence of the relaxation rate is expected to follow the conventional Redfield relation:

$$\lambda(H_{LF}) = \lambda_0 + \frac{2\gamma_\mu^2 \langle H_{loc}^2 \rangle \nu}{\nu^2 + \gamma_\mu^2 H_{LF}^2} \quad \text{S (7)}$$

where $\gamma_\mu$ is the muon gyromagnetic ratio, $\langle H_{loc}^2 \rangle$ is the local field distribution, and $\nu$ is the spin fluctuation rate. As shown in Figure S5 (b), the experimental $\lambda(H_{LF})$ data measured at 0.1 K deviate from the Redfield behaviour. Instead, the data follow a sublinear power-law dependence $\lambda \propto H_{LF}^{-\gamma}$ with $\gamma \approx 0.7$, where $\gamma$ is a fitting exponent, indicating non-exponential spin correlations. In addition, a time–field scaling analysis of the LF-μSR spectra reveals partial collapse of the asymmetry curves when plotted as $A(t, H) = A(t/H^\gamma)$ (Figure S5 (a)), further supporting unconventional spin dynamics in $CaCu_3Ir_4O_{12}$.

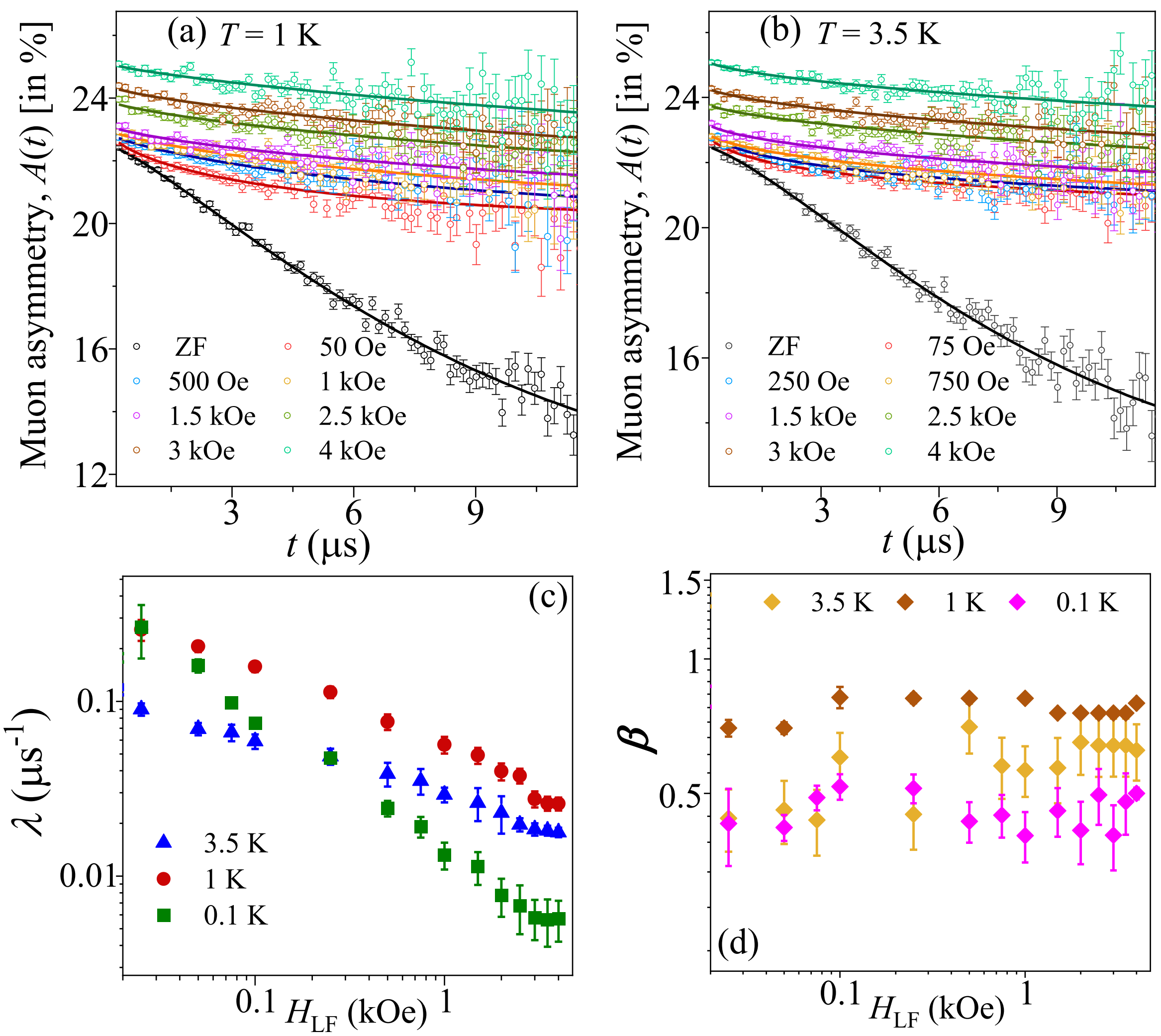

**Fig S4.** (a) Time evolution of the longitudinal-field (LF)-μSR asymmetry spectra at $T$ = 1 K (a) and 3.5 K (b) along with the fittings (solid-coloured lines) for some selected LFs. Comparative LF-dependence of the muon-spin relaxation rate, $\lambda$ (c) and the stretched exponent, $\beta$ (d) between three different temperatures ($T$ = 0.1, 1, 3.5 K) on a log-log scale.